\documentclass{elsart}
\usepackage{graphicx,harvard,amssymb}

\def\vrho{\varrho}
\def\dm{{\rm CDM}}
\def\gal{{\rm gal}}
\def\gas{{\rm gas}}
\def\o{{\rm o}}
\def\scale{{\rm scale}}

\def\inf{{\rm in}}

\def\ltsim{\raise 2pt \hbox {$<$} \kern-1.1em \lower 4pt \hbox {$\sim$}}
\def\ltapprox{\raise 2pt \hbox {$<$} \kern-1.1em \lower 5pt \hbox {$\approx$}}
\def\gtsim{\raise 2pt \hbox {$>$} \kern-1.1em \lower 4pt \hbox {$\sim$}}
\def\gtapprox{\raise 2pt \hbox {$>$} \kern-1.1em \lower 5pt \hbox {$\approx$}}

\def\arcmin{$^{\prime}$}

\def\Ang{$A^{\circ}$}

\def\phat{\hat{p}}

\def\eg{{\it e.g.,~}}
\def\ie{{\it i.e.,~}}

\def\hhi{~$h_{50}^{-1}$}

\def\aa{A\&A}

\def\aas{A \& A Sup. }
\def\apj{ApJ}
\def\aj{AJ }
\def\apjl{ApJ Lett. }
\def\apjs{ApJ Sup. }
\def\mnras{MNRAS}

\def\nat{Nature }
\def\na{NewA}
\def\cpc{{\rm Comp.\ Phys.\ Comm.\ }}

\makeatletter
\def\elsartstyle{%
	\def\normalsize{\@setfontsize\normalsize\@xiipt{14.5}}
	\def\small{\@setfontsize\small\@xipt{13.6}}
	\let\footnotesize=\small
	\def\large{\@setfontsize\large\@xivpt{18}}
	\def\Large{\@setfontsize\Large\@xviipt{22}}
	\skip\@mpfootins = 18\p@ \@plus 2\p@
	\normalsize
}
\makeatother

\def\bibcode#1{}
\def\astrobj#1{#1}
\def\url#1{{\ttfamily\def\/{/\discretionary{}{}{}}#1}}

\newcommand{\href}[2]{#2 (\texttt{#1})}

\begin{document}

\begin{frontmatter}
\title{Evidence for
 Shock Acceleration and Intergalactic
 Magnetic Fields in a Large-Scale Filament of Galaxies ZwCl 2341.1+0000}
\author[iucaa]{Joydeep Bagchi \thanksref{email} \thanksref{*author}} 
\author[garching]{Torsten A. En{\ss}lin \thanksref{email}}
\author[garching]{Francesco Miniati \thanksref{email}}
\author[upso]{Stalin C.S. \thanksref{email}}
\author[upso]{M. Singh \thanksref{email}}
\author[birmingham]{Somak Raychaudhury \thanksref{email}}
\author[chennai]{Humeshkar N.B.  \thanksref{email} \thanksref{now}}

\address[iucaa]{Inter-University Centre for Astronomy and Astrophysics (IUCAA), 
     Post Bag 4, Ganeshkhind, Pune 411007, India}
\address[garching]{Max-Planck-Institut f{\"u}r Astrophysik, Karl-Schwarzschild-Str.~1, 
     85740 Garching, Germany}
\address[upso]{State Observatory, Manora Peak, Naini Tal 263129, India}
\address[birmingham]{School of Physics \& Astronomy, University of Birmingham, 
Edgbaston, Birmingham B15 2TT, UK}

\address[chennai]{
Dept. of Physics, Loyola College, University of Madras, Chennai 600034, India}

\thanks[email]{E-mail addresses: joydeep@iucaa.ernet.in (Bagchi), \\
ensslin@mpa-garching.mpg.de (En{\ss}lin), \\
fm@mpa-garching.mpg.de (Miniati), \\ stalin@upso.ernet.in
(Stalin), msingh@upso.ernet.in (Singh), \\
somak@star.sr.bham.ac.uk (Raychaudhury), deephum@yahoo.com (Humeshkar).}
\thanks[now]{Present address: School of Physics, University of 
Hyderabad, Hyderabad 500046, India.}
\thanks[*author]{Corresponding author. Fax: +91-20-5690760. 
e-mail: joydeep@iucaa.ernet.in}

\begin{abstract}
We report the discovery of large-scale diffuse radio emission from
what appears to be a large-scale filamentary network of galaxies in
the region of cluster \astrobj{ZwCl 2341.1+0000}, and stretching over an area of
at least $6\,h^{-1}_{50}$ Mpc in diameter. Multicolour CCD
observations yield photometric redshifts indicating that a significant
fraction of the optical galaxies in this region is at a redshift of
$z\!=\!0.3$. This is supported by spectroscopic measurements of 4
galaxies in the Sloan Digital Sky Survey (SDSS) at a mean $z\!=\!0.27$. We present VLA
images at $\lambda=$20 cm (NVSS) and 90 cm, showing the detailed radio
structure of the filaments.  Comparison with the high resolution FIRST
radio survey shows that the diffuse emission is not due to known
individual point sources.  The diffuse radio-emission has a spectral
index $\alpha \lesssim -0.5$, and is most likely  synchrotron
emission from relativistic charged particles in an inter-galactic
magnetic field. Furthermore, this optical/radio structure is detected
in X-rays by the ROSAT all-sky survey. It has a 0.1--2.4 keV luminosity
of about $10^{44}$ erg~s$^{-1}$ and shows an extended highly non-relaxed
morphology. These observations suggest that \astrobj{ZwCl 2341.1+0000} is
possibly a proto-cluster of galaxies in which we are witnessing 
the  process of structure
formation.  We show that the energetics of accretion shocks
generated in forming large-scale structures are sufficient to produce
enough high energy cosmic-ray (CR) electrons required to explain the
observed radio emission, provided a magnetic field of strength $B \gtrsim
0.3 \mu$G is present there.  The latter is only a lower limit and the
actual magnetic field is likely to be higher depending on the
morphology of the emitting region.  Finally, we show results from a
numerical simulation of large-scale structure formation including
acceleration of CR electrons at cosmological shocks and magnetic field
evolution. Our results are in accord with the observed radio
synchrotron and X-ray thermal bremsstrahlung fluxes. Thus we conclude
that the reported radio detection is the first evidence of cosmic-ray
particle acceleration taking place at cosmic shocks in a magnetized
inter-galactic medium over scales of $\gtrsim 5\,h^{-1}_{50}$ Mpc.

\end{abstract}

\end{frontmatter}

\section{Introduction}
\label{intro}

Recent advances in observational cosmology have revealed that the
large-scale distribution of galaxies in the Universe has a
honeycomb-like structure, where the principal morphological elements
are interconnected networks of large filaments and sheets of galaxies
surrounding huge regions almost devoid of galaxies (the `voids')
\cite{Einasto97,Bond96,Doroshk96}. The enduring quest has been to
understand how such diverse structures emerge out of primeval density
fluctuations that grew over time due to the effects of gravity.

An important role in the structure-formation process is played by the
large-scale shocks that form as the primordial density fluctuations
become non-linear and the accretion flows on collapsing structures
become supersonic
\cite{1998ApJ...502..518Q,Miniati00,Ensslinetal00}.
Likely these shocks are responsible
for heating of most of the diffuse intergalactic medium
(IGM) up to $\approx 10^{5-7}$K \cite{CenOstr99}. Given their large
sizes and long lifetimes, these shocks have also been proposed to be
the sites for the acceleration of very high energy 
cosmic-rays up to $10^{18}- 10^{19}$ eV
\cite{Norman95,Kang96}. In addition, it has been recently pointed out
that the cosmic-ray ions accelerated at intergalactic shocks
could accumulate in the formed structure, storing a significant fraction
of the total energy there \cite{mrkj01}. Exploration of such ideas through
direct observations is truly important because even after close to a century since
cosmic-rays were discovered by Victor Hess in 1912,
we do not know how and where they are accelerated. 

Direct evidence for the ability of cosmic shock waves to
accelerate particles is given by the observed association of the
so called `cluster radio relic' sources with locations where
shock waves are expected from X-ray observations \cite{Ensslin98}.
Diffusive shock acceleration may be operative at
these locations and responsible for the radio emitting electrons.
But also shock re-illumination of fossil radio plasma in remnants
of former active radio galaxies is a viable explanation
\cite{Ensslin01}. The latter mechanism seems to
be the explanation of at least some of the cluster radio relics,
since in some cases very filamentary and torus-like morphologies
could be resolved. Such morphologies are predicted by numerical
simulations of the shock passage of a fossil radio plasma bubble
\cite{EnsslinBrueggen02}. However it is quite possible that
some of the cluster radio relic sources are caused by diffusive
shock acceleration of thermal electrons \cite{mjkr01}.                       

In order for particles to be accelerated, magnetic fields have to be
present at the shock waves. Magnetic fields are observed in
the intra-cluster medium (ICM) of clusters of galaxies by Faraday
rotation of polarized background sources \cite{2001ApJ...547L.111C}.
They are further revealed by
the presence of cluster-wide Mpc scale radio halos in some clusters of galaxies,
believed to be the result of synchrotron emission of relativistic
electrons accelerated in magnetic fields. The role of the magnetic fields
in clusters is still a matter of debate, but some recent works indicate
important dynamical influences in the ICM
\cite{2001ApJ...547L.111C,2001ApJ...549L..47V}. Outside clusters, no
firm detections of magnetic field in the intergalactic medium (IGM)
has yet been reported to our knowledge, but speculations
on the magnitude of magnetic fields in IGM range from $10^{-7}$G to
$10^{-12}$G.

The origin of cosmic magnetic fields is currently unknown and for this
reason any observational evidence of them  in the IGM   
environment is of great importance.
Magnetic fields could have a primordial origin or they could
have been seeded and amplified during the relatively recent history of large
scale structure formation. In this latter scenario, the Biermann
battery mechanism \cite{Biermann51} operating at either large-scale
shocks \cite{Kulsrud97,Ryu98} or ionization fronts \cite{gnedin00} has been
proposed as a viable model for generating seeds of strength
$10^{-19}-10^{-23}$G, to be subsequently amplified by turbulent
motions and/or galactic dynamos up to $\mu$G level. Alternatively, the
origin of cosmic magnetic fields has been attributed to the pollution
of the inter-galactic medium by winds and outflows from primeval
galaxies \cite{kronberg99} and/or quasars and radio galaxies
\cite{furlanetto01,krishna_wiita01}. In most of these
cases, magnetic fields would be expected to be present not only inside
galaxy clusters, but also in filaments.             
There were earlier attempts to
estimate the magnetization of the Universe from quasars
\cite{1990ApJ...364..451D,2001APh....16...47M}. 

The detection of magnetic fields in the IGM outside clusters of
galaxies would help the understanding of the origin of magnetic fields
within galaxies and galaxy clusters.  At the same time, observational
evidence for cosmic-rays in the intergalactic medium would allow us to
further our understanding of high energy phenomena occurring in cosmic
structures which have recently received much attention [see \eg
\citeasnoun{1996IAUS..175..333F} and references therein].
Observationally exploring these issues would provide important clues
for many physical problems, including, among others, the theories of
ultra-high energy cosmic-ray acceleration and their intergalactic
propagation, the origin of radio and hard X-ray emission from galaxy
clusters, and the contribution to the diffuse gamma-ray
background from shock-accelerated TeV cosmic-ray
electrons \cite{Loeb2000}.

In this paper we report on the observational evidence of large-scale 
diffuse radio synchrotron emission (\S~\ref{radobs})
over an extended region around \astrobj{ZwCl 2341.1+0000}, which is likely
to be a proto-cluster of galaxies containing several large 
filaments (\S~\ref{optobs}). The ROSAT X-ray observation of this structure 
is presented in \S~\ref{rosat}. The result of a search for 
association of
\astrobj{ZwCl 2341.1+0000} with other galaxy 
concentrations on a super-cluster scale is reported in \S~\ref{superscale}. 
Our observations 
imply the existence of magnetic fields and relativistic electrons at GeV
energies over scales of several Mpc (\S~\ref{radobs} \& \S~\ref{anlytic}). 
The observed radio emission is
compared with theoretical predictions from analytical (\S~\ref{anlytic}) and numerical
modeling (\S~\ref{numerical}) of particle acceleration at structure formation shocks. 
Finally in \S~\ref{conclude} we discuss the implications of our results and
the outlook for future work.

Unless stated otherwise, we adopt an $\Omega_{m}=1$ Einstein-de Sitter
cosmology and the Hubble constant $H_{\rm 0} = 50\, h_{\rm 50}\,\,{\rm
km\,s^{-1}\,Mpc^{-1}}$. The angular scale is $330 \, h^{-1}_{50} \rm
kpc\ $ per arcminute at a redshift z=0.3.  The radio spectral index 
$\alpha$ is defined such that flux density $F_{\nu}$ is a  power-law
$F_{\nu} \, \propto \nu^{\alpha}$.

\section{The \astrobj{ZwCl 2341.1+0000}}

The multi-Mpc scale filamentary network of galaxies in \astrobj{ZwCl
2341.1+0000} was discovered in an ongoing program to search for the
large-scale diffuse radio emission ($\sim 1 \, \rm Mpc$ or larger)
originating in distant clusters of galaxies. The primary aim of this
program is to investigate the origin and evolution of high energy
relativistic particles and magnetic fields in the IGM of large cosmic
structures. To this end, we searched for the presence of diffuse radio
emission in the field of several Abell and Zwicky clusters or groups
with measured/estimated redshifts $\gtrsim 0.15$. Although not a
statistically complete sample, it nevertheless contains many clusters
that have no detailed information available so far.  As a first step,
we searched for radio emission in the `New VLA All Sky Survey'
(\href{http://www.cv.nrao.edu/nvss/}{NVSS}) images at the
1.4 GHz frequency \cite{Condon98}. The NVSS survey uses the VLA's most
compact D-array configuration and is well suited for detection of
large structures of angular scales up to $\sim 15$ arcmin in extent. The 
completeness limit for point sources is about 2.5 mJy and the angular
resolution (beam size) is about 45 seconds of arc.
 
In the NVSS data, we discovered, among others, a large-scale diffuse
emission feature in the region of galaxy concentration
\astrobj{ZwCl 2341.1+0000} located at R.A.  $23^h 43^m 39^s.7$
Dec. $+00^\circ 16^\prime 39^{\prime\prime}$. It is classified as a distant
group 5 or ``extremely distant'' cluster with cluster diameter 22\arcmin \
by \citeasnoun{Zwicky61}). Further
literature and data search  with the NASA Extragalactic Database
(\href{http://nedwww.ipac.caltech.edu}{NED}) showed no
additional information on this concentration of galaxies. However, the
optical images of the Second Generation Palomar Digitized Sky Survey
(\href{http://archive.eso.org/dss/dss}{DSS-2})
revealed a very interesting cosmological scale (projected size
$\sim 25$ arcmin or $8 h^{-1}_{50} \rm Mpc\ $) filamentary network
of faint galaxies, which had the distinctive non-relaxed morphology of
a forming structure. Motivated by these findings, and with an aim to
investigating the physical properties of this object in greater detail,
we obtained several more detailed observations at radio and optical
wavelengths.  These observations and the results of their analysis
form the subject matter of the following sections. 

\section{Optical data} \label{optobs}

\subsection{CCD imaging and analysis} 

We obtained several deep optical and NIR images with the 1-m
Carl-Zeiss reflector at the  State Observatory, Nainital,
India, in November 1999. Due to constraints of time and scheduling, only
data in Johnson $V$, $R$,  and $I$ filters were obtained using a CCD camera
equipped  with a  2K $\times$ 2K pixels chip, giving a plate scale
of $0.37$ arcsec/pixel, and a field of view of about 13 arcmin at the f/13
Cassegrain focus. The gain and the read-out noise were 10 $e^-$/ADU
and 13.7 $e^-$ respectively. The sky conditions during the observations
were photometric but the  seeing was moderate at $\sim$2 arcsec FWHM. We
took several frames in each filter with total exposure time sufficient
for adequate photometric accuracy. Photometric zero points and colour
transformations were defined by observing Landolt standards.  The
initial optical processing of the CCD frames was done using the IRAF
software and the final object detection, magnitude evaluation and object
(star/galaxy)  classification was performed using the `SExtractor' image
analysis program \cite{Bertin96}. 

The resulting source catalogue contains a total of 271 galaxies
detected to a magnitude limit of $m_{V} = 21.9$, $m_{R} = 20.9$, and
$m_{I} = 20.2$, and which were detected independently in all three passbands. The
total magnitudes for galaxies at the faint end had typical errors
of $\sim$0.05--0.10 mag. The extinction corrections are small
($A_B<0.1$ mag over the area of observation) due to the high galactic
latitude ($b=-58^\circ$) position.  The parameters for the extinction were
obtained from the IRAS based galactic extinction estimates by
\citeasnoun{Schlegel98}. The astrometric conversion from the pixel to
equatorial co-ordinates was performed with IRAF employing several
bright astrometric stars identified on the CCD images.

\subsection{Optical morphology and luminosity}\label{morphlum}

The Fig.~\ref{fig1} shows the {\it R}-band co-added image of the galaxy
concentration.  Morphologically it is best described as a long S-shaped
main filament of galaxies extending over 12 arcmin ($\sim 4 h^{-1}_{50}$
Mpc, which is the size of the CCD frame), and branching 
structures of several Mpc-scale galaxy
sub-filaments mainly to the east and north-east of the main chain. The
photometric redshift estimate places the structure at a redshift of
z$\approx 0.3$, supported by spectroscopic evidence from the SDSS
survey (see \S \ref{sdss} below).

If the entire structure is actually a web-like formation at a single
redshift and not an artifact of projection (evidence in support of
this is presented below), the implications are clearly very
important. The branching chain structure and multiple peaks in the
galaxy distribution might indicate hierarchical merging of cosmic
structures distributed on Mpc scales, at a moderately high
redshift. This type of structure is to be expected in evolving
Universe according to the ``cosmic web'' theory of structure formation
\cite{Bond96}, and it is frequently seen in computer simulations such
as the the Hubble-volume N-body simulation \cite{colberg2000}, and also in
observational surveys such as the Las Campanas  redshift survey
\cite{Doroshk96}.  These results predict that matter in the Universe
should be concentrated along filaments, and clusters of galaxies
should be found where these filaments intersect.  It is interesting to
note that the filamentary morphology of \astrobj{ZwCl 2341.1+0000} is
quite similar to a more distant proto-cluster \astrobj{RX
J1716.6+6708} at $z\!=\!0.81$ \cite{Henry97}.

The average integrated galaxy $V$ luminosity of 12 randomly selected
positions on the main filament and side chains, after correcting for the
local background offset from main structure, yields a value of $2.7
\times 10^{11} L_{\odot}/{\rm arcmin^2} \ (\pm 54 \%)$ for the observed average
luminosity of the filament. The indicated dispersion about the mean value
 represents  the fluctuation of the surface brightness
occuring over the galaxy filaments, rather than the uncertainty of measurement.

\subsection{Galaxy colours \& estimating photometric redshifts} \label{photomet}

We estimated photometric redshifts for the galaxies in our catalogue
to ascertain the mean redshift of the concentration, and to verify
that most of the galaxies projected on the sky plane near this concentration
represent an association in real space. This method is a powerful tool
for detecting high spatial density regions such as clusters and groups
and has recently been applied to the study of galaxy populations in
high-z clusters of galaxies \cite{Lubin00,Gladders98}. From our
photometric catalogue of galaxies in $V$, $R$ and $I$ bands, a set of 6
plots; three of colour-magnitude (C-M) diagrams and three of
colour-colour (C-C) diagrams, were generated. Figs. \ref{fig2} \&
\ref{fig3} show the $(V-I)$ and $(R-I)$ colours vs. $I$ magnitudes
for all the galaxies.  On both planes, a
well defined linear sequence of early-type (E/S0) galaxies is
clearly visible. 

Early-type galaxies form the dominant population in
cores of all clusters, rich or poor, and across a wide range of
redshifts, low to $z > 1$ (e.g., \citeasnoun{Gladders98}). Their
ubiquitous colour-magnitude sequence, first noted locally by
Baum \cite{Baum59}, is now well documented in many clusters, ranging
from \astrobj{Coma} to clusters at redshifts up to $z\! =\!  1$
\cite{Stanford98,Ellis97,Gladders00}. The colours of elliptical
galaxies become bluer as they become less luminous. We have detected
this red C-M sequence on all three C-M diagrams, which shows that at
least the dominant galactic population consists of early-type galaxies
that are part of this filament at a single redshift. This is a good
indicator of association of galaxies in a large-scale structure, 
as a random projection of field
galaxies can not generate the red sequence which is characteristic of
clusters. On the C-M diagrams, apart from E/S0s, a population of blue
galaxies (late spirals and irregulars) can also be seen in the
lower-right region.

From the position of each galaxy in a 3-dimensional colour space,
consisting of orthogonal axes $(V-R)$, $(V-I)$, and $(R-I)$, 
the three colour-colour diagrams were obtained. These diagrams
can be used to identify the galaxies of various Hubble types and to
obtain the redshifts. The Fig.~\ref{fig4} and Fig.~\ref{fig5} show two
of these projections that show galaxies of very wide range of colours. A
distinct cluster of data points near $(V-R)=0.9$, $(V-I)=1.6$ and
$(R-I)=0.7$ can be seen on these plots.  These identify a coeval
population of red galaxies of early Hubble types at a single
redshift. Therefore, we have further proof that \astrobj{ZwCl
2341.1+0000} is indeed a real concentration of galaxies.

The properties of the galactic populations in these filaments were
investigated using the spectrophotometric evolutionary models of
\citeasnoun{Guiderdoni88}. These models give evolutionary synthetic
spectra for a number of spectral classes reproducing the range of
spectrophotometric properties for the Hubble sequence.  From this work,
the data on predictions for apparent magnitudes and colours of  galaxies
of wide range of redshifts were obtained and fitted to the observed
colours and magnitudes. The evolutionary tracks, representing the
redshift evolution of galaxy colours, were plotted on the colour-colour
diagrams. The data included the effects of K-correction, evolutionary
correction and of the emission of nebular component  plus the correction
for internal extinction (see \citeasnoun{Guiderdoni88} for exact
details).  The colours of E and S0 galaxies are useful in determination
of redshift because of the uniformity of their spectral properties  and
because of a large 4000 \Ang \, break due to the Balmer edge in their
spectra provides the strongest signal for photometric redshift estimation.
In Fig.~\ref{fig4} and Fig.~\ref{fig5} it is apparent that the
evolutionary tracks for the early type E/S0 galaxies fall closest to
the observed data  cluster at redshift $z\!\sim\! 0.3$. The colours for
E/S0's at redshifts $z\!=\!0.2$ or $z\!=\!0.4$ both showed a much poorer
fit to the data, particularly on the $(V-I)$ vs. $(V-R)$ plot which has the
steeper slope. 

To avoid projection effects, the actual fit of the spectrophometric
model colours to the observed data was performed in a 3-dimensional
colour space. At redshift $z\!=\!0.3$, the model colours for the E/S0
galaxies are: $(V-R)_{mod}=0.92$, $(V-I)_{mod}=1.58$, and
$(R-I)_{mod}=0.66$. Enclosed within a sphere of radius 0.30 colour
index units centered on the model point, we found an agglomerate  of 79 data
points. These are the likely to be the E/S0 galaxies (having similar
colours) that have high probability to belong to the filament. 
The radius of 0.30 colour-index units 
takes into account the intrinsic spread in colours of E/S0
galaxies observed in clusters. The plotted position of these galaxies
on the C-M diagrams (Fig.~\ref{fig2} and Fig.~\ref{fig3}) forms well
defined linear sequences characteristic of early type galaxies in
clusters, confirming that they are E/S0 galaxies.  The mean and
standard deviation of these galaxy colours were calculated as:
$(V-I)_{E/S0}=1.59\pm 0.12$, $(V-R)_{E/S0}=0.88 \pm 0.09$ , and
$(R-I)_{E/S0}=0.72\pm 0.11$. Clearly, the model for E/S0 galaxies fits
the cluster photometric observations quite well (within $\lesssim 0.5
\sigma$). From this evidence it was estimated that the actual redshift
is $z\!=\!0.30 \pm 0.05$, which was used in all our analysis and
interpretation. A more sophisticated analysis aiming at greater
precision was not attempted as the accuracy of the redshift estimate
was limited by the availability of only three photometric colours. The
accuracy of the photometrically estimated redshift is sufficient for
the purpose of the present paper. In addition, the plotted template
colours for the galaxies of Hubble types Sa, Sb, Sc, Sd and Irr at
z$\!=\! 0.3$ fitted the data well, as can be seen in Fig.~\ref{fig4}
and Fig.~\ref{fig5}, providing independent verification of the
redshift estimate.

\subsection{Spectroscopic redshifts} \label{sdss}

The current release of the ongoing
Sloan Digitized Sky Survey \cite{york:2000} includes
the spectra of four of the galaxies in our catalogue, from
which redshifts are shown in Table~\ref{tab:sdss}. The location of 
these
galaxies are shown  by arrows in Fig.~\ref{fig1}.  The mean
redshift from these four galaxies, $z\sim$0.267, is consistent
with our photometric redshift estimate in the previous section.
For the rest of paper, we will assume the redshift of the filamentary structure
to be $z\!=\! 0.3$.

\begin{table}[htb!]
\caption{\bf Spectroscopic redshifts from the SDSS\label{tab:sdss}}
\begin{flushleft}
\begin{tabular}{lllll}
\hline
Galaxy& R.A.& Dec.& Magnitude & Redshift \\
no.& J(2000)& J(2000)& & \\
\hline
1&  $23^h43^m38.2^s$&$00^\circ 19^\prime 46^{\prime\prime}$& 18.9&0.272093 \\
2& 23~43~40.7\ & 00~20~30\ &                               17.6&0.260870 \\
3&   23~43~34.6\ & 00~20~37\ &                               18.0&0.269087 \\
4&   23~43~21.7\ & 00~19~33\ &                               17.7&0.266947 \\
\hline
\end{tabular}
\end{flushleft}
\end{table}


\section{The radio observations} \label{radobs}

\subsection{The low resolution VLA (NVSS) radio data at 20 cm \label{NVSS}}

The 1.4~GHz ($\lambda=$20~cm) radio map obtained from the 
\href{http://www.cv.nrao.edu/nvss/}{NVSS} archives is shown
superposed  on the optical $R$-band CCD image in Fig. \ref{fig6}. The
original data of 45~arcsec (FWHM) resolution was convolved with a
Gaussian to obtain a 60~arcsec (FWHM)  
resolution beam in order to better detect
the large-scale diffuse emission regions. The $1 \sigma$ noise level on
this image is within range $\approx 0.5-0.6$ mJy/beam, whereas on the unsmoothed
original image it was $\approx 0.45$ mJy/beam.  

Morphologically, the source appears to have a complex structure, but mainly 
consisting of two parts: the southern
section  (at R.A. $23^h 43^m 49^s$, Dec. $00^\circ 14^\prime 
14^{\prime\prime}$) is $\approx$ 5~arcmin ($1.65
\, h^{-1}_{50} \rm Mpc\ $) in size and is entirely diffuse in nature, and the
peak emission here is about 4 mJy/beam. 
The northern section (at R.A. $23^h 43^m 39^s$, 
Dec. $00^\circ 18^\prime 44^{\prime\prime}$),  of similar
$\approx$5 arcmin angular size,  shows evidence of both diffuse and localized
radio emissions. Within the northern radio structure, the data allows
to resolve two more radio components which are henceforth  called for convenience 
NN (the northern peak of 6 mJy/beam)
and NS (the southern peak of 11.5 mJy/beam). The integrated flux density, position
and spectral index
for  various regions of the source are reported in Table~\ref{tab:radio}. The
spectral indices were calculated by convolving the NVSS 
radio map to the broader resolution
of the 320 MHz radio map discussed below.
A  faint `radio-bridge' type extension 
at R.A. $23^h 43^m 43^s$, Dec.
$00^\circ 16^\prime 31^{\prime\prime}$ between 
the two sections is an interesting feature. The surface 
brightness of this `radio-bridge' is about 1.5-2.5 mJy/beam and therefore it is
detected at the level of about 3-4 $\sigma$. A strong curvature 
in the main filament of galaxies can be clearly seen in this region. However, the
curved section is not located on the radio-bridge 
but shifted about 1.5 arcmin to the NE of it. We note that this pattern is again
reproduced in the ROSAT X-ray map (Fig.~\ref{fig11}) 
which is discussed in better detail in \S~\ref{rosat}.  

More significantly, the entire structure visible at radio wavelengths is of very
large extent (at least 10 arcmin or $3.3 \, h^{-1}_{50} \rm Mpc\ $) and appears to be
roughly  aligned with the main optical galaxy filament which is of similar
size. Thus the radio and optical structures are possibly related
and co-spatial. We show below, from high resolution radio data, that 
only one  elliptical galaxy could be associated with  a radio peak. The
rest of the radio emission therefore possibly originates from diffuse
synchrotron radiation (as inferred from the radio spectrum) 
from the inter-galactic medium. 

\begin{table}[htb!]
\caption{\bf Radio data\label{tab:radio}}
\smallskip
\begin{flushleft}
\begin{tabular}{lllllll}
\hline
 Region& R.A.& Dec.& $S_{320}$& $S_{1400}$& Ang. size& Sp. indx.\\
                  & J(2000)& J(2000)& mJy & mJy& (arcmin)&
$\alpha_{320}^{1400}$ \\
\hline
  North half(total)$^{a}$& --& --& $50\pm10$& $28\pm2$& $\sim 5$& $-0.40\pm0.20$\\
NN$^{a}$& 23 43 39.9& 00 20 53& $22\pm5$& $11\pm2$& $\sim 2$& $-0.47\pm0.20$\\
NS$^{a}$& 23 43 39.4& 00 18 41& $28\pm5$& $17\pm2$& $\sim 2$& $-0.34\pm0.15$\\
  South half$^{a}$& $\sim$23 43 49& 00 14 14& $42\pm8$& $20\pm2$& $\sim
5$& $-0.50\pm0.15$\\
  Extended 90 cm$^{b}$& $\sim$23 44 03& 00 22 00&
$36\pm10$& $< 10$& $\sim 4$& $<-0.9$\\
  Total source& --& --& $128\pm16$& $48\pm5$& $\sim 12^{c}$& \\
\hline
  Radio Power$^{d}$& --& --& $ 5.6\pm 0.7$ & $2.1 \pm 0.2 $ & --& -- \\
\hline
\end{tabular}
\end{flushleft}
\par\noindent
{\em } a: As defined in the VLA 1.4 GHz observations.
b: The excess diffuse emission found with VLA 320 MHz data.
c: The largest angular size at 320 MHz. The size measured along
the `ridge-line' is about 18.4 arcmin.
d: The total monochromatic radio power in units of $10^{25} \,
h_{50}^{-2}$ W Hz$^{-1}.$
\end{table}

\subsection{High resolution VLA `FIRST' image and optical identification
of possible point sources} 
\label{first}

The  VLA low resolution 20 and 90 cm 
radio images discussed here have only the modest 
resolutions of  $\sim$1-1.8 arcmin,
while in a portion of  sky covered by a galaxy cluster there can
be several hundred galaxies. It is therefore possible that a
significant part of the radio emission could arise from the blend of
radio sources in the cluster (and distant background point sources). It is
therefore very important that we should identify the regions of genuine
large-scale diffuse emission from superposed point sources. 
To check
for this serious source of confusion, we have cross correlated the
optical \astrobj{ZwCl 2341.1+0000} cluster field with the NRAO  
\href{http://sundog.stsci.edu/first/description.html}{`FIRST'} radio
catalogue. The `FIRST' survey, conducted with the VLA B-configuration,
contains high
resolution ($\approx$ 5 arcsec beam) radio maps at 1.4 GHz with a flux
density threshold of 1 mJy and a typical rms of 0.15 mJy/beam. 
The astrometric reference frame of the maps
is accurate to 0.05 arcsec and individual sources have $90\%$ confidence
error circles of radius $< 0.5$ arcsec at the 3 mJy level. To achieve an
equivalent degree of precision, for optical identification we have used the 
Palomar Digitised Sky Survey (DSS-2) red sensitive plate for which  the
astrometric accuracy is better than $\pm0.5$ arcsec r.m.s. With such 
good radio and optical astrometric precision, it is possible to identify
superposed point sources, if any are present. We note that the large-scale
structures $\gtrsim$ 30 arcsec angular size are resolved out by 
the FIRST survey, but compact features, such
as unresolved radio-cores or jets, can be easily identified.

Fig.~\ref{fig7} shows the FIRST image (contours) superposed on the
DSS-2 image (gray scale) for optical identification of radio sources. 
The image covers the northern section
of the radio filament detected by the NVSS. FIRST survey finds radio
counterparts at the position of the two northern sources NN and NS (see
\S~\ref{NVSS} on the NVSS data), but the southern region of diffuse emission
is completely resolved out and shows no compact radio peak.
Similarly, the extended large-scale  diffuse emission
detected in the 320 MHz data (discussed below) 
also does not show any point components in the FIRST
data. The integrated flux densities of the two sources detected by FIRST are: in
the northern source NN $1.8\pm0.4$ mJy and in the 
southern source NS $9.3\pm0.6$ mJy. The integrated
flux densities of the same two sources mapped by NVSS are: in 
the northern source NN $11\pm2$ mJy and in the southern source NS $17\pm2$ mJy 
(see Table~\ref{tab:radio}). Clearly, this indicates that while 
FIRST is actually able to map
the regions close to peak emission in both theses sources, there is considerable
amount of diffuse emission associated with both which is largly resolved out
due to lack of short interferometric spacings (as expected of VLA B-array data).
We note that even with this limitation the FIRST map of the 
southern radio source NS shows a  diffuse morphology and some hints of
partially resolved internal structure 
consisting  possibly of two parallel filaments $\approx$ 6 arcsec apart. 
These features can be better seen
on the detailed  radio image  shown in Fig.~\ref{fig8}. We
need more sensitive radio maps to better understand what these features really are.
 
No object classified as a galaxy with the 
CCD data could be located on the stronger southern radio peak NS. A faint
$m_{v}=20.81$ galaxy (G1) was located 6.2~arcsec to the north of the radio
peak (Fig.~\ref{fig7}), and another brighter $m_{v}=18.21$ galaxy (G2) 
6.5~arcsec to the south-east direction, but their identification with
the radio source is doubtful. The combined radio and optical positional
errors correspond to an error circle of radius 1 arcsec and both galaxies are
shifted by $\approx$\ 6$\sigma$s from the radio peak. The  photometric 
colours  of galaxy G1 are quite red and not consistent with those of a
standard elliptical galaxy at $z\!=\!0.3$, but are more like those of a
higher redshift galaxy. Thus based on  high resolution radio and optical data
we can not find any good optical identification for this source. Most likely the
diffuse source NS is part of the overall large-scale radio filament.  

Near the weaker northern radio source NN, an E/S0
galaxy (G3) of magnitude $m_{v}=18.19$ could be located 
at R.A. $23^h~43^m~40^s.4$, Dec. $+00^\circ~20^\prime~54^{\prime\prime}$,
poitioned only 1.7~arcsec east from the FIRST radio peak.  Due to its close
proximity and consistent photometric colours (for redshift z$\approx 0.3$),
we consider it to be a
possible candidate host galaxy/AGN for the origin of northern radio
peak. At present the true relationship of this weak radio peak with the rest of the
extended diffuse emission found with NVSS 
and 320 MHz data is not very clear. The NVSS
map shows an extension of about 2 arcmin size to the west of the radio peak, whereas
the 320 MHz map shows both this extension as well as another `plume' to the
east of radio peak NN. We consider two possibilities:  first, that the optical
galaxy G3 (with a weak radio nucleus) is just superposed on the rest of the radio
filament which is not directly related to this galaxy. This possibility can not
be ruled out from the present data.  
Second, that this
galaxy is the source of all the radio emission in the region of the source NN. If
this is true then can this radio source be a `head-tail' type radio galaxy commonly
found in clusters ? This interpretation is not fully supported  by  radio
spectral index $\alpha_{320}^{1400}=-0.47\pm0.20$ as reported in 
Table~\ref{tab:radio}. The spectral index is quite flat for a head-tail radio source
which generally have their integrated spectral 
indices quite steep $\lesssim \ -0.7 $. On the other
hand the
fairly flat spectral index of -0.47 is consistent with the similar flat radio 
spectra
noted over the other regions of the radio filament (Table~\ref{tab:radio}). In
that case the radio spectral evidence suggests that  both the 
northern sources NN and NS could
be  local regions of stronger emission embedded within the large radio filament. 
The cause could be that the physical conditions differ from place to place within the
filament, with local effects on particle acceleration which then is reflected in the
intensity of radio emission.

Apart from this possible radio source, no other optical galaxy could be
identified with  any part of the extended diffuse radio emission
mapped by the VLA at 20 and 90 cm wavelengths. Therefore, given the evidence at
hand, we consider it to be
unlikely that a major part of the large-scale radio structure is due to
the blend of radio-galaxies in the cluster or in the background. It is
very likely that the radio emission originates entirely in the diffuse
intra-cluster medium and derives its power from an energetic process
spread across a region several Mpc wide.

\subsection{The meter-wavelength 320 MHz VLA image and spectral
indices} \label{90cm}

The field of \astrobj{ZwCl 2341.1+0000} was further observed by the VLA
on May 12, 2000 at  327~MHz ($\lambda=$\, 90~cm) for a total on-source
duration of 2.2~hr.
The compact C-array observation was obtained with an aim to achieving
resolution and large-scale structural mapping sensitivity comparable to
the NVSS data. The two IFs (intermediate frequencies) 
were centered at 327.5 MHz (IF1) 
and 321.5 MHz (IF2) with
bandwidths of 3.12 MHz in each IF. The data were obtained in spectral-line
mode and amplitude (flux density) 
calibration was obtained by boot-strapping to \astrobj{3C 48} which
was assumed to have a flux density of 42.547 Jy and 42.966 Jy at 
IF 1 and 2 respectively. The 327.5 MHz data was found to be 
heavily contaminated by radio interference and therefore could not be used. The
quality of 321.5 MHz data was found to be satisfactory and it was used for
further analysis. For analysis we used the NRAO `AIPS' package. 

Low-frequency radio imaging of faint sources 
is always challenging due to large primary beam
area of  antennae, resulting in detection of many strong sources over
this area. In addition, the effects of sky-curvature can not be neglected in
mapping and cleaning of fields measuring several degrees in size. In order 
to mitigate these problems as much as possible, we used the task `IMAGR' that
permits the `faceted' wide-field mapping 
and cleaning option. We mapped upto 53 small
fields (facets) centered on strong sources located {\it a priori}
within the $\rm 5.6 \times 5.6 \ deg^{2}$ field of view
centered on \astrobj{ZwCl 2341.1+0000}, and then simultaneously cleaned the entire set.
At the end of cleaning cycle the clean components  from all facets together were
restored back onto a single field to generate a large wide-field image. 
We point out that due to the use of separate tangent point for each
field, 3-D imaging,
and the subtraction of clean 
components from un-gridded data (preventing aliasing of side-lobes), the quality
of the final clean image  is fairly good. In order to emphasise the
large-scale features and supress the oscillations in the side-lobes, the
visibility data were tapered by a  Gaussian  that had $30\%$ weight
at the 2 k$\lambda$
length of the interferometric spacing. This has resulted in better detection 
of faint diffuse features, but at the
cost of angular resolution which is now 108 arcsec FWHM (circular Gaussian) 
on the radio
map shown in Fig.~\ref{fig9} . The  level of  background  noise 
on the cleaned image is about 2.5 mJy/beam r.m.s. as against
1.3 mJy/beam expected from pure thermal noise. This about factor 2 higher
noise is mainly the effect of residual low-level side-lobes of 
several strong sources of 0.2 Jy-1.5 Jy flux density occuring nearby that
limit the  dynamic range achievable to $\approx$ 400. 

In order to facilitate  a comparison of
the  radio maps at 320 MHz and  NVSS map at 1400 MHz, we also
show two wide-field images covering identical $\rm 1.5 \times 1.5 \ deg^{2}$
fields. The contour levels shown are in 
multiples of $\sim 1\sigma$ noise on each
map. For a better comparison, the original NVSS map has been
convolved with a Gaussian to obtain a broader final beam size of $90 \times 90$
$\rm arcsec^{2}$. These images can be accessed on-line by linking to   
\href{http://www.iucaa.ernet.in/~joydeep/vla/}{this address}.
 
Comparison of  90~cm radio map (Fig.~\ref{fig9})
with the 20~cm map (Fig.~\ref{fig6})
reveals the detection of all the features present on the 20~cm map, albeit 
at a coarser resolution. Furthermore, starting from radio source NN
(in the northern half of 20 cm image) and extending $\approx$ 5 arcmin
to its east, an excess region of  
diffuse emission can be detected at 90~cm. This can be found at
R.A. $\approx \, 23^h~44^m~03^s$, and
Dec. $\approx \, 00^\circ 22^\prime 00^{\prime\prime}$. Although the reality
of features detected both at 20~cm and 90~cm are not in any doubt, 
we nevertheless 
introduce a caveat on the excess diffuse radio feature.
Although the $\sim 1 \, h^{-1}_{50} \rm Mpc\ $ sized
`plume' to the east of radio peak NN does suggest some connection, but it is
quite faint and  detected at the significance 
level of about 2$\sigma$ (surface brightness 
$\approx$ 5 mJy/b)
only. Therefore it must be treated with due caution.
The two radio peaks of diffuse emisiion,
further towards east, that also seem to be linked with the plume, are somewhat better
detected at the levels of 3$\sigma$ (peak 7.7 mJy/b) and 5.6$\sigma$ 
(peak 14 mJy/b). But due to faint nature of 
these radio structures, they
should also be treated with caution  until confirmed with stronger signal in
an improved observation. In case  these features are real, they would make the total
radio size much larger than what is visible at 20 cm wavelength (see below).  

The  optical morphology of galaxies underlying the 90 cm extended 
structure is shown in Fig.
\ref{fig10}, where the radio contours are shown superposed on the
Palomar DSS-2 red sensitive photograph. From this image, and
also from the higher
resolution `FIRST' radio data discussed above, no galaxy was found to
be apparently associated with any region of the diffuse radio structure. 

The largest observed angular size at 90 cm is  about 12 arcmin, which
corresponds to a projected dimension  of $\approx 4 \, h^{-1}_{50}$ Mpc.
The size when measured along the spine
of the filamentary structure was found to be about
$6 \, h^{-1}_{50}$ Mpc, thus making it comparable or larger than some
of the largest known radio structures in the Universe:
the giant radio galaxies \astrobj{3C236} ($5.8 \, h^{-1}_{50}$
Mpc) and \astrobj{NVSS 2146+82} ($3.9 \, h^{-1}_{50}$ Mpc)
\cite{Strom80,Palma00}, and also the largest known radio-loud quasar
\astrobj{HE 1127-1304} ($2.4 \, h^{-1}_{50}$ Mpc;
\citeasnoun{Bhatnagar98}). However, there is no evidence available to show
that the radio emission in any way constitutes a giant radio galaxy
(see \S~\ref{grgs} below). A peculiar radio morphology and the pronounced
lack of symmetry in the observed structure are particularly noticeable in
this source.

The extended region was detected at a significance level of 2--5$\sigma$.
Therefore, for the extended part, the radio spectral
index is steep ($\alpha <-0.9 $), estimated from its non-detection
at 20~cm to the flux density limit of 10~mJy.  Estimates of  flux
densities and the  spectral indices for the various regions of the radio
structure are listed in Table~\ref{tab:radio}.
The overall negative spectral indices
in all regions ($\alpha \lesssim -0.5$) indicate a non-thermal   mechanism for
radio emission, most likely the  synchrotron radiation of relativistic
charges accelerated in a magnetic field. Therefore it is necessary that both the
particles and fields, distributed on similar spatial scales, should be
present to generate the radio emission. {\it To our knowledge, this is the
first observational evidence of distribution of 
particles and magnetic fields on spatial scales
$\gtrsim \ 5 \ h^{-1}_{50}$ Mpc}.

\subsection{On the possibility of radio source in \astrobj{ZwCl 2341.1+0000}
being a giant radio-galaxy} \label{grgs} 

The so-called giant radio-galaxies (GRGs) are a rare class of largest
radio sources whose projected linear size between the radio lobes is
$\gtrsim 1$ Mpc. Their other physical characteristics are that they are
mostly bi-symmetric, have edge-brightened FR-II (Fanaroff-Riley class
II) radio morphology. Their radio power is in the transition zone
between FR-I (low-powered systems) and FR-II (high-powered ones).
Furthermore, they tend to occur in regions of both low galaxy density and low
intergalactic matter density. In view of the extremely large size noted
for the radio source in \astrobj{ZwCl 2341.1+0000}, we ask-- can it be another
GRG?

We have several pieces of evidence which seem to rule out this
possibility. First, the radio morphology is very peculiar, lacking any
semblance of bi-modal or other symmetry or identifiable radio-jets or
large-scale radio-lobes as expected of GRGs. Second, on the high
resolution `FIRST'  map we do not find any evidence of a compact radio
core associated with a centrally located optical galaxy  which could be
assumed to be the source of the entire radio structure. Third, the
projected linear size in the range 4--6 Mpc is very large even by the
standards of GRGs. \citeasnoun{Schoenmakers01}, in their  systematic
study of several GRGs found that beyond the linear size of 2 Mpc the
number of GRGs cuts-off sharply and sources above this size are
extremely rare. There appears to be a strong luminosity evolution in the
GRG population, such that in the majority of them the activity in the
central engine is extinguished over the time it takes for the
radio-lobes to expand to size $\gtrsim 2$ Mpc. Therefore, the extreme size
of the present radio source and its morphological peculiarities  suggest
that it is not powered by the energy outpouring in the expanding radio
lobes. Some other energy generation mechanism spread over the entire radio
structure appears to be at work here. We show below that possibly this is
diffusive shock acceleration in structure formation flows. 
It is therefore unlikely that \astrobj{ZwCl 2341.1+0000} is a giant
radio-galaxy.

\section{X-ray detection in the ROSAT All-sky survey } \label{rosat}

An X-ray counterpart of the collapsing filament \astrobj{ZwCl
2341.1+0000} can be found in the ROSAT all-sky survey 
(\href{http://www.xray.mpe.mpg.de/cgi-bin/rosat/rosat-survey/}).  Almost the
entire sky was surveyed by the PSPC-C instrument on ROSAT in scanning
mode over six months in 1990, such that every point in the sky was
observed with an effective exposure time of between 117 and 443.9
seconds.

Fig.~\ref{fig11} shows the RASS observation, smoothed with a Gaussian
of $\sigma_r\! =\! 4$ pixel $\sim$ 1 arcmin, as contours superposed on
the optical (Palomar DSS-2 red) image.  We have used only the 0.5-2~keV band
of the observations to produce the image, since the softer band shows
no significant signal in this region (higher background). The r.m.s. level of
 background noise on the smoothed image is 
typically $5\times10^{-3}$ counts/s,
which was estimated over several nearby regions. The first contour
is shown at the level of about 4 $\sigma$ ($21\times10^{-3}$ counts/s) and
next contours are spaced linearly in steps of 1 $\sigma$. 
The highest
X-ray peak ($45.9\times10^{-3}$ counts/s, $9\sigma$ detection)
in this region is at R.A. $23^h 43^m 39.6^s$, Dec
$+00^\circ 19^\prime 40.3^{\prime\prime}$ (J2000).  
The secondary peak ($37.7\times10^{-3}$ counts/s, $7.5\sigma$ detection)
is at R.A. $23^h 43^m 44.4^s$, Dec $+00^\circ 15^\prime
08^{\prime\prime}$.  
The positional accuracy of the X-ray data was checked with respect to
the optically identified point X-ray sources in the field and was 
found to be $\approx$ 15 arcsec as expected of
the RASS \cite{Voges99}.  

The net flux in the 0.1--2.4 keV band in the region shown in
Fig.~\ref{fig11} is $2.52 \times 10^{-13}$ erg cm$^{-2}$~s$^{-1}$ 
($\pm$43\%), corresponding to a $3\sigma$ detection.  A galactic neutral
hydrogen column of $3.7 \times 10^{20}$ cm$^{-2}$ was assumed
\cite{lockman90}, and no $k$-correction was applied ($< \ 10\%$). We assumed a
gas temperature of $T\!=\! 5$ keV in the calculation.  
At redshift
$z\!=\!0.3$, this flux corresponds to bremsstrahlung 
X-ray luminosity of  $1.1 \times 10^{44} \,
h_{50}^{-2}$  
erg~s$^{-1}$. The largest angular size of the detected X-ray structure is
about 8 arcmin ($2.6 \, h^{-1}_{50}$ Mpc), comparable to the size of the
radio structure. The structure of the X-ray source appears quite
non-relaxed,  which suggests that
this system is possibly undergoing a violent gravitational
merger of subclusters. Therefore the ROSAT data 
supports the possibility that  \astrobj{ZwCl 2341.1+0000} is in fact
a systen undergoing structure formation. This is  a likely scenario in view of the
filamentary morphology of galaxies noted in optical.

Although the RASS detection is not  strong 
enough in order to provide firm inferences, we
nevertheless show a spatial comparison between the radio and X-ray emitting plasma by
superposing the 327 MHz radio image and the X-ray contours in Fig.~\ref{fig12}. Here we
can locate the stronger northern X-ray structure in the northern part of the radio
filament while the secondary X-ray peak to the south is in the vicinity of the southern
diffuse radio structure and both seem to be joined by an X-ray filament/bridge of 
$\sim$ 2.2 arcmin (730 $h_{50}^{-1}$ kpc) size 
($\approx 27\times10^{-3}$ counts/s or $5.4\sigma$ detection). We note a strong
curvature in the main filament of galaxies located about 1.5 arcmin to the east
of this X-ray  bridge (Fig.~\ref{fig11}). This pattern is also visible on the
20 cm radio and optical superposition shown in Fig.~\ref{fig6}.
Another extended X-ray filament/plume
of $\sim$ 3 arcmin ( 900 $h_{50}^{-1}$ kpc)  
size, and detected at the $4-6 \ \sigma$ level, 
can be found to the east of the northern X-ray peak where several smaller filaments
join with the larger main filament of galaxies.
Interestingly, both the X-ray peaks as well as the filaments are
apparently displaced from the nearby radio peaks (by as much as 1 to 1.5 arcmin -
significantly larger than the X-ray positional error). While this clearly indicates
that the X-ray emission mechanism can not be the non-thermal 
inverse Compton radiation, however it can provide a 
diagnostic tool to probe the physics of the multi-phase (thermal and non-thermal)
intra-cluster medium in comparison with the structure formation simulation results.
A similar strong
mismatch between the locations of relativistic
and thermal plasma is noticeable in the numerical simulation carried out by
us in \S~\ref{numerical}. We  defer the discussion on it's physical significance to
\S~\ref{numerical} where it will be dealt with  
in comparison with the theoretical
model.

\section{The large-scale environment} \label{superscale}

The linear dimensions of the optical and radio structures described
here, although quite large, are both restricted by the maximum possible areas that
could be imaged with 
the optical and radio telescopes used. We have no reason to believe
that these are not parts of a larger super-structure, which needs to be
verified for its important implications.  As of now, neither deep
wide-field optical images nor extensive redshift measurements are
available in order to check whether or not this concentration of galaxies
visible in the region of \astrobj{ZwCl 2341.1+0000} is part of a larger 
supercluster ($\sim$10--100 Mpc) scale structure. 

As a step in this direction,  we have used the Palomar DSS-2 survey
images and the NASA Extragalactic Database
\href{http://nedwww.ipac.caltech.edu}{NED} to search  for
other known clusters and groups within a radius of 2~deg ($40 \,
h^{-1}_{50}$ Mpc) of the center. There are eight identified galaxy
concentrations which are listed in Table~\ref{tab:lss}. 
The two nearest clusters,
\astrobj{Abell 2644} and \astrobj{RXC J2341.1+0018}, are at much lower
redshifts to be associated with \astrobj{ZwCl 2341.1+0000}. 
The other five Zwicky clusters, situated
within 40--75 arcmin, are of comparable richness, but their redshifts
are not known.  However,  there is a very rich cluster \astrobj{Abell
2631} (richness class 3, $z\!=\!0.2730$), which is only 1.5~degree ($30
\, h^{-1}_{50}$ Mpc) away on the sky  and at a comparable redshift,
making it a  possible association on a  super-cluster scale.  Deep
wide-field imaging and spectroscopy of a large area in the
\astrobj{ZwCl 2341.1+0000}  region could reveal 
the true supercluster-scale environment
of this large forming structure.

\begin{table}[htbp!]
\caption{\bf Large scale environment: known groups and clusters 
within 2 degrees ($40 \, h^{-1}_{50} \rm Mpc\ $) \label{tab:lss}}
\smallskip
\begin{flushleft}
\begin{tabular}{lllll}
\hline
  Cluster or group& R.A.& Dec.& Redshift&  Separation\\
                  & J(2000)& J(2000)& & (arcmin)\\ 
\hline 
  \astrobj{ZwCl 2341.1+0000}& $23^h43^m39.7^s$& 
           $+00^\circ 16^\prime 39^{\prime\prime}$ &\llap{$\sim$}0.30& 0.0\\
  \astrobj{RXC J2341.1+0018}& 23~41~06.3\ & +00~18~53\ & 0.110& 38.4\\
  \astrobj{Abell 2644}      & 23~41~09.8\ & +00~05~38\ & 0.069& 39.1\\
  \astrobj{ZwCl 2338.3-0022}& 23~40~51.8\ &--00~05~22\ & ...& 47.4\\
  \astrobj{ZwCl 2343.9+0029}& 23~46~27.7\ & +00~45~40\ & ...& 51.0 \\
  \astrobj{ZwCl 2342.2+0049}& 23~44~45.7\ & +01~05~40\ & ...& 51.7\\
  \astrobj{ZwCl 2344.9-0025}& 23~47~27.8\ &--00~08~20\ & ...& 62.2 \\
  \astrobj{ZwCl 2345.7+0030}& 23~48~15.7\ & +00~46~41\ & ...& 75.2 \\
  \astrobj{Abell 2631}      & 23~37~39.7\ & +00~17~37\ & 0.273& 90.0 \\       
\hline 
\end{tabular}
\end{flushleft}
\end{table}

\section{Evidence of particle acceleration in 
shock waves associated with large-scale structure formation} \label{anlytic}

\subsection{The necessity of a large-scale particle acceleration process} \label{formula}

The radiative life-time ($t_{\rm life}$) of a relativistic electron in a
weak magnetic field ($ \sim 10^{-7}$ G) is dominated by inverse
Compton loss (IC), and it is given by
\begin{equation}
 \rm t_{life} = 80.36 \times 10^{6} \ yr  \ 
\left( {\gamma \over 10^{4}} \right)^{-1} = 
  4.1 \times 10^{7} \ yr \ 
  \left( {E_{e} \over { GeV}} \right)^{-1}, 
\end{equation}
where $\rm E_{e}$ is the electron energy, $\gamma$ is the electron
Lorentz factor and  the redshift is assumed to be $z\!=\!0.3$. 
At the strongly turbulent  magnetic field environment near the sites of particle
acceleration (e.g., shocks), the diffusion  may be governed by the Bohm
approximation, in which the diffusion coefficient is (e.g., \citeasnoun{Drury83})
\begin{equation}
 \rm \kappa_{\rm Bohm} = 3.31 \times 10^{23} \ \left( {E_{e}
\over { GeV}} \right) 
   \left( {B \over {10^{-7} \ G}} \right)^{-1} \ cm^{2} \ s^{-1}. 
\end{equation}
Therefore, the diffusion length-scale ($\rm L_{diff}$) for an electron
within the inverse Compton radiative cooling life-time $\rm t_{life}$
is
\begin{equation}
 \rm L_{diff} = ( \kappa_{\rm Bohm} \cdot t_{life})^{1/2} = 
 6.7 \ \left({B \over {10^{-7} \ G}} \right)^{-\frac{1}{2}} \ pc. 
\end{equation}
The linear size of the observed radio emission $ \rm L_{radio}$ is
about several Mpc which is much bigger than $L_{diff}$. Therefore, electrons
once accelerated in a localized point source such as a radio galaxy or AGN
and diffusively transported cannot be the source of the large-scale
radio emission. The discrepancy between the Bohm diffusion length-scale
and the radio structure size is so large, that even in the case of
ordered magnetic fields, where diffusion should be more effective
(than in the Bohm approximation), electrons are still unable to cross
the emission region within a radiative life-time.

Furthermore, we have not found any evidence of radio jets originating
at a central nucleus and transporting bulk kinetic energy to large
distances in expanding radio-lobes. Therefore, it is required that the
particles (electrons) must be accelerated {\it in situ}, and the
acceleration mechanism should be such that it is extended over scales
of several Mpc, and long-lived over the dynamical time-scale $\rm
t_{dyn}\sim 10^{9}$ y. In addition, they are required to be energetic
enough to rapidly achieve relativistic energies on an acceleration
time-scale $t_{acc}$ such that $\rm t_{acc} < t_{life} < t_{dyn}$. We
show below that the diffusive shock acceleration associated with the
formation of large-scale structure   \astrobj{ZwCl 2341.1+0000}
is the most viable
mechanism in operation.

\subsection{Energetics of shock acceleration and estimates
of the magnetic field strengths involved \label{energmag}}

The extremely large linear size for the radio structure, its peculiar
radio morphology unlike any radio galaxy, and its association with
this striking optical filament of galaxies, suggest to us that we are
possibly witnessing a new and hitherto unknown type of radio
phenomenon. The NVSS VLA map has a central surface brightness of
$S_{\rm 1.4GHz} \approx 5.5 \, {\rm mJy / arcmin^2}$ for the diffuse
emission of the source \astrobj{ZwCl 2341.1+0000} (1 mJy = $10^{-29} \
\rm{W\,Hz^{-1}\,m^{-2}}$). From radio data, a power-law
radio-frequency spectrum with spectral index $\alpha \sim -0.5$ is found
in the diffuse parts. Assuming this to be the spectral index 
across the radio frequency range 
$\nu_1 = 10$ MHz to $\nu_2 = 10$ GHz, the  representative radio surface-brightness 
of this structure can be estimated to be $Q_{\rm radio} = 1.6 \times
10^{42}\,{\rm erg\, s^{-1}\, Mpc^{-2}}$. Any possible acceleration mechanism should
be energetic enough in producing this order of radio brightness over the entire
filament.

The mechanism for the radio emission is very likely to be the
synchrotron radiation of relativistic charges accelerated in a
magnetic field, and therefore the presence of a magnetic field energy
is necessary.  To obtain an estimate of the field strength in the
observed structure, we minimize the total non-thermal energy density,
required to produce the observed radio emission, with respect to the
magnetic field and write for the minimum energy field [which is close to the
equipartition field, \cite{Miley80}], 
\begin{equation}
 \rm B_{me} = 0.35\,
\left[\eta \ sin^{3/2}\phi \ \left({d\over {\rm
Mpc}}\right)\right]^{-{2 \over 7}}\,\mu G.
\end{equation}  
Here, we have tacitly assumed an energy ratio of protons to electrons
equal to unity, and $\alpha=-0.5$.  For a range of volume filling
factor $\eta$= 1--0.5, and the angle of field lines relative to line
of sight $\phi = 90$--45 deg, our estimated field spans the range $\rm
B_{me} =$ 0.5--0.3 $\mu$G, assuming a cylindrical geometry with
observed length $= 3.3$ Mpc (10 arcmin), width $= 1$ Mpc (3 arcmin),
and assumed depth $d= 1$ Mpc, consistent with the morphology of the
filament as observed at 20 cm. The calculated minimum energy
field $B_{me}$  depends weakly on source geometry, 
the nonthermal spectral index, and the
frequency range of the synchrotron radiation, usually taken to be 0.01-100 GHz.

It is more likely, however, that the emission comes from a thin layer
close to the accretion shock surface. This implies a much smaller
filling factor.  If we assume that $\eta = 0.01$ and $\phi = 90$--45 deg,
the magnetic field obtained is in range $\rm B_{me} = 
1.3-1.5 \, \mu$G. The
radiative age of the 1.4 GHz radio-emitting electrons, undergoing
inverse Compton and synchrotron emission losses, is then only $\lesssim
50\,{\rm Myr}$. 

The diffuse radio emission from a region several Mpc across can only
originate in a powerful energy source of a similar size. To our
knowledge, only the expected accretion flow of inter-galactic matter
onto the filament is sufficiently extended, long-lived, and energetic
to overcome the radiation losses of the relativistic electrons over a
Mpc-scale filament. The exact physical mechanism of the electron
acceleration is yet to be revealed. It is logical to assume, from its
filamentary structure and high redshift, that \astrobj{ZwCl 2341.1+0000}  is
evolving towards, but is still far from becoming, a centrally
condensed galaxy cluster. The accretion shocks associated to forming structures
are typically strong, with Mach numbers, $M \sim 5-10$, as opposed
to the weaker merger shocks which typically have $M \sim 2$ \cite{Miniati00}. 
This makes them ideal sites for high energy
particle production. Diffuse regions of large-scale radio emission,
commonly known as `radio relics' and `radio halos', and the detection
of temperature structure in several nearby galaxy clusters are
possibly related to such  shocks
\cite{willson70,Kim89,1996IAUS..175..333F,1997A&A...321...55D,Bagchi_mnras98,Limaneto_etal01,Ensslin98,1999NewA....4..141G,Ensslin01,mjkr01,EnsslinBrueggen02}.
However, no such evidences have hitherto been found in supercluster-scale
filamentary structures at high redshifts.

We propose here that the large-scale radio emission and magnetic field
detected in \astrobj{ZwCl 2341.1+0000}  are strong evidences for 
diffusive shock-acceleration \cite{Bell78,Blandford78}
of particles in structure formation flows \cite{Ensslin98,mjkr01}. A
natural consequence of this process would be a power-law spectrum of
the momentum distribution. The acceleration process giving rise to a
pool of high-energy cosmic-ray particles (both electrons and protons), 
and specifically the radio emission
tracing the radiation from energetic electrons interacting with the
magnetic fields. The inverse Compton scatter of cosmic microwave
background photons from accelerated electrons may also generate a
detectable hard X-ray flux from the structure (e.g., \citeasnoun{Bagchi_mnras98}). 

We test the viability of the suggested mechanism by showing that 
the available kinetic power in
the accretion flows in this galaxy filament is sufficient to explain
the observed radio luminosity and the unobserved inverse Compton
radiation. For this, we consider a simplified model of the filament. The
main structure is described as an infinite extended, linear,
self-gravitating, isothermal cold dark matter (CDM) filament.  The
radial profile of the dark matter mass distribution is
\[  \vrho_\dm(r) = \vrho_{\dm, \o}\,
\left[ 1+{r^2 \over 8\, r_\scale^2} \right]^{-2}, \]  
where r is the radius,
and 
$r_\scale = \sigma_\dm/(4 \,\pi\, G \,\vrho_{\dm, \o})^{1/2}$
is the typical scale of the dark matter distribution which has a central
density of $\vrho_{\dm, \o}$ and a 1-dim. velocity dispersion of
$\sigma_\dm$. $G$ is the gravitational constant.

The observed galaxies can be regarded as test particles since they do
not significantly
affect the gravitational potential of the dark matter. 
Since the galaxies 
are expected to have a smaller velocity dispersion $\sigma_\gal$
than the dark matter, we assume a conservative velocity bias of $b_{\rm
v} = \sigma_\gal / \sigma_\dm \approx 0.8$ \cite{1990ApJ...352L..29C}.
This leads to a spatially more concentrated profile for the galaxies  
\[
\vrho_\gal(r) = \vrho_{\gal, \o} \,  
\left[ 1+{r^2 \over 8\, r_\scale^2} \right]^{-2/b_{\rm v}^2}. 
\]  
We estimate that the visible diameter of the filament of roughly $D_\gal
= 100$ arcsec corresponds to the region where the projected galaxy
density is above 30\% of the central value, which implies $D_\gal
\approx 4\, r_\scale$.  As shown in \S~\ref{morphlum}, the
observed average $V$-band luminosity of the filament of galaxies is
estimated to be $2.7 \times 10^{11} L_{\odot}/{\rm arcmin^2}$.  For a
central mass to light ratio of $M/L_V = 100$, this translates into a
central dark matter mass density of $ \vrho_{\dm,\o} = 4.5\times
10^{-26}\, {\rm g\, cm^{-3}} \,\cos \theta$. Here, $\theta $ is the
(unknown) angle of the filament axis and the normal to sky-plane, which we
assume in the following to be $\theta = 45^\circ$.  The velocity
dispersion of the dark matter can then be estimated to be $\sigma_\dm
\approx {690}$ km/s.

The estimated high central density and high velocity dispersion of
this galaxy filament are comparable to that of a small galaxy
cluster. The velocity of the in-falling matter should reach a final
velocity of $v_\inf = \sqrt{3}\, \sigma_\dm = 1200$  km/s, in
order to provide the velocity dispersion of the filament. This
estimate of high velocity of infall is in accord with previous
estimates based on hydrodynamic simulations
(e.g., \citeasnoun{RyuKang97}), and with those derived theoretically
(e.g., \citeasnoun{Bertsch85}). The characteristic incident velocities
are $u_{s} \approx 10^{3} \, \rm km\ s^{-1} \ [(M_{cl}/R_{cl})/(4
\times 10^{14} M_{\odot}/Mpc)]^{1/2}$ in accretion shocks around
clusters of galaxies of given mass-to-radius ratio $M_{cl}/R_{cl}$.

We now estimate the rate of dissipation of kinetic energy by the
accretion shock over the surface area of the filament, given by $Q_{\rm
kin} = \frac{1}{2} \,\vrho_{\gas,\inf}\, v_\inf^3$. We assume the
infalling low density baryonic gas 
to have a proton number density $n_{\rm P}=
10^{-5} \,{\rm cm^{-3}}$, corresponding to the mass density of
$\vrho_{\gas,\inf}\ =1.67 \times 10^{-29} \,{\rm g \ cm^{-3}}$ (i.e., 
$\vrho_{\gas,\inf}/
\vrho_{crit}\ = 3.55$). This
gives $Q_{\rm kin} = 1.4 \times 10^{44} \,{\rm erg\,s^{-1}\,
Mpc^{-2}}$, which is sufficient to provide an integrated 
(over the radio spectrum)  radio brightness $Q_{\rm
radio} \approx 1.6 \times 10^{42}\,{\rm erg\, s^{-1}\, Mpc^{-2}}$ even if
this kinetic power is converted to radiative energy with an efficiency
of $\approx 1\%$.  

In addition, this kinetic power allows for the low general efficiency
of the diffusive shock-acceleration process ($\sim 10 \%$) and 
cooling due to inverse Compton radiation $Q_{\rm IC} = 10.5 \ Q_{\rm
radio} \ ( B/ \mu \rm G)^{-2}\,(1+z)^{4}$. 
Therefore, balancing the
energy gain by $Q_{\rm kin}$ with the losses in the energy-budget, we
obtain an upper limit to the inverse Compton loss $Q_{\rm IC} < 1.4
\times 10^{44}\,{\rm erg\, s^{-1}\, Mpc^{-2}}$ and a lower limit to the
allowed magnetic field strength in the inter-galactic medium $B \gtrsim 0.56
\, \mu \rm G$. 
We note that even lower values of the
magnetic field strength are allowed if the available kinetic power is
higher than estimated here, either due to a higher infalling mass
density or a larger velocity of the accretion flow.

Following   \citeasnoun{Drury83}, 
the mean acceleration time scale for electrons
to reach a given energy is
\begin{equation}
 \rm t_{acc} = {8 \over u_{s}^2} \ \kappa_{\rm Bohm} = 8.36 \ 
\left( {E_{e} \over { GeV}} \right) \
\left( {B \over {10^{-7} \ G}} \right)^{-1} \ 
\left( {u_{s} \over {10^{3} \ km \ s^{-1}}}\right)^{-2} \ yr, 
\end{equation}
where $ \rm u_{s}$ is the shock speed. Here we have taken the limit of
strong shock, so the downstream speed $\rm u_{2}$ is related to upstream
speed  $\rm u_{1}$ by $\rm u_{2}=u_{1}/4$, and $\rm \kappa/u=constant$. For
the diffusion coefficient, once more the Bohm diffusion condition
$\kappa=\kappa_{\rm Bohm}$ is invoked. Clearly, $\rm t_{acc} \ll t_{life}$, so in
comparison to the radiative loss time scale, the acceleration process is
virtually instantaneous. Also, the acceleration time-scale is much shorter
compared to the expected life time of 
shocks $\rm t_{acc} \ll t_{shock}(\sim 10^{9} y)$.
Under these
conditions,   
the maximum energy attained by  electrons in
the acceleration process can be calculated by setting the acceleration
and cooling times equal ($\rm t_{acc}=t_{life}$), which gives
\begin{equation}
\rm E_{e,max} \approx  2.2  \ \left( {B \over {10^{-7} \ G}} 
       \right)^{1/2} \ \left(
{u_{s} \over {10^{3} \ km \ s^{-1}}}\right) \ TeV. 
\end{equation}
Protons are expected to achieve even higher energies due to their far less radiative
losses.
Thus the cosmological accretion shocks in  large-scale forming
structures such as  \astrobj{ZwCl 2341.1+0000} are indeed capable of 
accelerating particles to  very high energies. 

\section{Comparison with a numerical model \label{numerical}}
\subsection{Description of simulations}

In order to investigate the  nature of  radio emission and magnetic field
presented in \S~\ref{radobs} \& \S~\ref{energmag}, we 
carried out a specific numerical
simulation of large-scale structure formation that, in addition to the
the dark matter and the baryonic components, also follows a passive
magnetic field and the cosmic-ray electrons (CREs). In the following we
briefly describe the technique employed for the treatment of each of
these components.


For the cosmological part we have used an Eulerian, grid based
Total-Variation-Diminishing hydro~+~N-body cosmology code
\cite{rokc93}. We have adopted a commonly favored $\Lambda$
+ cold dark matter model with the following parameters: total mass
density $\Omega_m=0.3$, vacuum energy density $\Omega_\Lambda=1-
\Omega_m=0.7$, baryonic mass fraction $\Omega_b = 0.04$, and normalized
Hubble constant $h_{50} = 1.34$ (i.e.,  
$\rm H_{\rm 0} = 67\, km\,s^{-1}\,Mpc^{-1}$). 
The initial density perturbations were distributed as a
Gaussian random field with a power spectrum characterized by a
spectral index $n$ = 1 and ``cluster''
normalization $\sigma_8 = 0.9$. The dark
matter component is described by $128^3$ particles, whereas the gas
component is evolved on a grid of $256^3$ cells. We chose a comoving
size for the simulation box $L_{\rm box}=100 h_{50}^{-1}$ Mpc. Such a size
should be large enough to allow the formation of groups and small
clusters of galaxies with temperatures up to a few keV.

With the above set up, the size of a computational cell corresponds to
about 400\hhi kpc, a dark matter particle mass to about 
$8.5\times 10^8$\hhi M$_\odot$,
and the spatial and mass resolution to a few times these values
respectively. The adopted resolution is somewhat modest.
However, it is higher than that 
characterizing the radio observations. In addition,
our tests indicate that coarse grid effects do not
affect the simulated electrons and their corresponding radio
emission too significantly, which are the primary objectives here.


The magnetic field is treated as a passive quantity, in the sense that
its dynamical role on the fluid behavior is completely neglected.  The
initial magnetic field is set to zero but magnetic field seeds are
generated at shocks in accord to the Biermann battery mechanism 
\cite{Biermann51}.  The
field strength is thereafter enhanced by shear flow through stretching
and by field compression \cite{Kulsrud97}. However, even at 
the end of the simulation,
the magnetic field within collapsed structures is well below
observational values [see \citeasnoun{Kulsrud97} for further details].
For this reason, the overall field strength normalization is arbitrary.
Nevertheless, the simulation will provide important information about
the topology and the relative strength of the magnetic field in
different regions of the flow.


Finally, the evolution of the CREs is computed by means of the
numerical code COSMOCR \cite{min01a,min02}.  The code is fully
described in the given references and here we shall only briefly
summarize the salient aspects of it.  We suppose that CREs are only
injected at shocks from the thermal plasma and we neglect the
contribution from any other source (e.g., secondary electrons produced
in inelastic p-p collisions, or injected from radio-galaxies).  The
injection at shocks is parametrized in the following way.  We compute
the fraction of injected cosmic-ray protons according to the `thermal
leakage' model [\eg \citeasnoun{kajo95}], and assume a fixed value
$R_{e/p}$ for the ratio of CREs to protons at relativistic energies.
This simplified approach, after \citeasnoun{elbeba00}, is motivated by
the fact that the physics underlying the injection at shocks of
primary electrons is very complex and not well understood.
Observational evidence suggests that this ratio is in the range
0.01-0.05 for the Galactic CRs \cite{muta87,mulletal95}.  With the
parameters assumed for our `thermal leakage' model, the fraction of the
thermal protons passing through shocks and injected as cosmic-rays is
$\sim 10^{-4}$ [see \citeasnoun{min01a} for details].

After their injection at shocks as described above, the CREs are
evolved taking into account both spatial transport and energy
losses/gains. This is achieved by solving numerically a Fokker-Planck
equation that has been integrated over finite momentum bins.
Basically, momentum space is divided into $N_p$ logarithmically
equidistant intervals, referred to as {\it momentum bins}.  The
electron distribution function $f(\hat{p})$, as a function of the
normalized momentum $\hat{p}\equiv p/m_ec$, in each spatial cell and
for each momentum bin is approximated by the following piece-wise
power law:
\begin{equation} \label{distf.eq}
f({\bf x}_i,\phat) = f_j({\bf x}_i) \, \phat^{-q_j({\bf x}_i)},
~~~~~~~~ 1<\phat_{j-1} \le \phat \le \phat_j,
\end{equation}
where $f_j({\bf x}_i)$ and $q_j({\bf x}_i)$ are the number normalization
and logarithmic slope
for a given cell and $\hat{p}$ bin.
With the above definition, the number density of particles
is given by $dN = 4\,\pi\,\hat{p}^2\, f(\hat{p}) d \hat{p}$.

Within each momentum bin $j$, and at each spatial grid point ${\bf x}_i$,
we follow the total number density and kinetic energy density 
of the CREs, defined as
\begin{eqnarray}
n({\bf x}_i,\phat_{j}) & = &  4\pi\;
\int_{\phat_{j}}^{\phat_{j+1}} f({\bf x}_i,\phat) \phat^2 d\phat \\
\varepsilon({\bf x}_i,\phat_{j}) & = &  4\pi\;
\int_{\phat_{j}}^{\phat_{j+1}} f({\bf x}_i,\phat) T(\phat) \phat^2 d\phat,
\end{eqnarray}
where $T(\phat)=(\gamma-1) m_ec^2$ is the relativistic
kinetic energy.
Further, for each momentum bin,
$q_j({\bf x}_i)$ is determined self-consistently
from the values of
$n({\bf x}_i,\phat_j)$ and $\varepsilon({\bf x}_i,\phat_j)$
defined above [see 
\citeasnoun{min01a} for details].
With this formalism, the evolution of $n({\bf x}_i,\phat_{j})$
in momentum space is described by the equation
\begin{equation}
\frac{\partial n({\bf x}_i,\phat_{j})}{\partial t}
= - {\bf \nabla\cdot [ u}\,n({\bf x}_i,\phat_{j})]
+ \left[
b(\phat) \;4\pi \;\phat^2\; f(\phat)\right] _{\phat_{j-1}}^{\phat_j}
+ Q({\bf x}_i,\phat_{j}),
\label{dce3.eq}
\end{equation}
where the first term on the right hand side describes advective
transport and $Q({\bf x}_i,\phat_{j}) $ represents a source term,
$i({\bf x}_i,\phat)$, integrated over the $j_{th}$ bin.  Finally,
$b(\phat)\equiv dp/dt$ describes mechanical and radiative loss terms
[see \citeasnoun{min01a} for further details].  For the energy range
of interest here, the most effective among these are synchrotron and
inverse Compton emission.

\subsection{Simulation results and comparison with observations}

The simulation data allow us to compute various quantities directly
related to the observations reported in \S~\ref{radobs}. Being
interested in radio emission, we have computed the synchrotron
emission from the simulated relativistic electrons and magnetic field
distribution. We have then integrated the emissivity along one of the
coordinate axis of the computational box to create a radio map. In
general we find that shocks around large, massive structures such as
groups and clusters, are relatively bright radio sources. A detailed,
quantitative analysis of the properties of radio emission at cosmic
shocks is presented in \citeasnoun{mjkr01}.

In order to investigate the physical conditions of the cosmic
environment where the CREs are accelerated, we also have obtained analogous
maps of inverse Compton hard X-ray (HXR) at 50 keV (from the same
relativistic electrons responsible for the radio emission) and of
X-ray emission in the 0.5-2 keV band from thermal bremsstrahlung of
the collapsed structures.

A portion of size 10 Mpc comoving corresponding to the same structure
from each of these maps is presented in Fig.~\ref{fig13}. Here we show
respectively radio emission in units of `$R_{e/p}^{-1}$ Jy pxl$^{-1}$' 
(top left panel), X-ray from thermal bremsstrahlung in units `erg s$^{-1}$
cm$^{-2}$ pxl$^{-1}$' (top right), and 50 keV HXR in units `$10^{-23}$
$R_{e/p}^{-1}$ erg s$^{-1}$ cm$^{-2}$ Hz$^{-1}$ pxl$^{-1}$' (bottom right).
In these radio/X-ray units, 1 pxl is about $\rm 0.34 \times 0.34 \, arcmin^{2}$. 
In the bottom left panel we also show 9 iso-contours of the radio emission,
ranging from $10^{-5.8}$ to $10^{-4.6}$ and separated by constant
factor of $10^{0.15}=\sqrt{2}$. We point out that the 
presence of $R_{e/p}$ in our radio and HXR units indicates explicitly
the dependence of the computed amount of non-thermal emission on the
assumed value for this parameter. 
In addition, the non-thermal emission in the reported figures 
depends on an assumed ionic energy injection efficiency of order $10\%$ 
and a magnetic field average strength of 1$\mu$G. 
Based on a comparison of the numerical and observational results,
below we indicate slightly modified values (for these quantities) 
that are required in order for us to reproduce the observed total 
radio emission.
Given the resolution of the simulation and the putative
redshift of the source, the original pixel size in the synthetic image
was $0.^\prime 34$. Thus to better compare with the observational radio
images, the value of the emission in the synthetic images has been
averaged over a region corresponding to $3 \times 3$ pixels. The
choice of this particular portion of the maps has the sole purpose of
showing the typical morphological and emission properties of a cosmic,
supersonic accretion flow. We briefly discuss similarities between 
the observed (Fig. \ref{fig6} \& Fig. \ref{fig9})
and simulated radio structures 
although we  {\it do not} expect matching between the two at any level.
We notice that although the contours in the synthetic radio
map appear more elongated than in Fig. \ref{fig6}, this is
particularly due to the use of much
fainter contours in the former image.

One feature of general interest concerns the relative 
location of the radio emission with respect to the thermal X-ray 
bright source, that is a group/cluster of galaxies.
In our synthetic radio map the brightest (southern)
emission region is located in correspondence of an accretion 
shock and sort of drapes around one side of an X-ray 
group/cluster. On the northern side of this X-ray structure
a fainter  extended radio source is also present.
Some of the emission there is  produced 
by acceleration regions associated to a smaller collapsing
object that happens to be on the same line of sight. Several more
smaller and fainter X-ray clumps are also present here.
In both cases however, one notes that the radio and thermal X-ray emitting 
regions do not overlap but are somewhat
displaced with respect to each other. 

In \S~\ref{rosat} we have shown how this pattern is also 
 present in the observational data and
becomes visible when a superposition of the radio emission at 
320 MHz (or 1.4 GHz) with the 
ROSAT X-ray map (Fig. \ref{fig11})
is carried out (Fig. \ref{fig12}). It is particularly evident 
in the southern section of the observational images where the
thermal X-ray peak lies shifted to the west from the diffuse
regions of radio emission which appears to surround the X-ray
region from north, east and south. The stronger X-ray peak to
the north as well as the extended filament to the south-east 
also do not have 
their exact radio correspondences, although  radio peaks are present
nearby within 1-5 arcmin. What physical process can produce such an
effect ? 

This is likely due to the fact that  the radio emission possibly
traces the peripheral regions of strong accretion 
shocks generating freshly accelerated
relativistic particles. These are the regions
at the  interface between the externel IGM and ICM.
In contrast, the bremsstrahlung
X-rays  come from interior regions of deep gravitational potential
wells in which shock heated and compressed gas is accumulating 
over cosmological
time scales. Note that we can confirm this picture from the synthetic 
images where we can easily delineate the regions of non-thermal 
and thermal emissions, and
non-thermal emission does correlate spatially with strong shocks.

The maps shown in Fig.~\ref{fig13} were used to compute the 
total flux from radio-synchrotron and thermal bremsstrahlung emission
to be compared with the observational values. Given the
radio and X-rays bremsstrahlung fluxes obtained from these plots, and
the dependence of the radio flux from the properties of the underlying
structure as derived in \citeasnoun{mjkr01}, we have estimated the physical
parameters characterizing our numerical model for the CR electrons and
the size of the underlying structure required to reproduce the
measured radio flux. In addition, using the bottom right panel of
Fig.~\ref{fig13}, we
have also computed the expected HXR flux between 20 and 80 keV.  In
Table~\ref{simul.tab} we report each of these quantities and we also
summarize the assumptions on shock acceleration efficiency and
magnetic field strength before we discuss them below.

\begin{table}[htb!]
\caption{\bf Simulation Results \label{simul.tab}}
\smallskip
\begin{flushleft}
\begin{tabular}{llllll}
\hline
 $P_{\rm CRE}/P_{\rm ram}$ &  $B$ & $S_{\rm sync}$(1.4 \ GHz) &  $F_{\rm HXR}$(20-80 \ keV) & $F_{\rm XBR}$(0.5-2.0 \ keV)  \\
               & $(\mu$G) & (mJy) & (erg s$^{-1}$cm$^{-2}$) & (erg s$^{-1}$cm$^{-2}$) \\
\hline
   $10^{-2}$ & 1.5 & 20-30 & $4\times 10^{-13}$ & 6$\times 10^{-13}$ \\
\hline
\end{tabular}
\end{flushleft}
\end{table}

The radio flux reported in the previous sections for 
\astrobj{ZwCl 2341.1+0000}  is quite large given the redshift of this
source. Nevertheless, 
according to the simulation results, a radio flux of a few $\times 10$ mJy
can be produced by shock accelerated CREs if about 1\% of the
shock ram pressure is converted into CREs and if the magnetic field
is of order of 1-2 $\mu$G. This computed flux accounts for the 
contribution from all the sources that are present in the 
synthetic image, although it is almost entirely produced
by the southern radio filamentary structure. 
The shock accelerated CREs have typically a flat distribution 
function, \ie $f(\phat) \propto \phat^{-q}$, $q\sim 4$, 
implying a radio spectral index $\alpha \approx -0.5$.
However, the total radio flux is produced by the contribution of 
various CREs distributions that, as they leave the acceleration 
region and propagate away from it, steepen due to inverse Compton and
synchrotron losses. Numerically, it is impossible to spatially
resolve these populations. However, it is possible to compute their
cumulative distribution function within the numerical cell. 
This is automatically given 
by the steady state solution of eq. \ref{dce3.eq} where the source
term describes the shock accelerated CREs and the losses are due 
to inverse Compton and synchrotron emission. This approximation
holds as long as the
CREs cooling time is much shorter than the  lifetime of shocks
(see \citeasnoun{min02} for further details). It is easy to show
that for a shock accelerated distribution function 
$f(\phat) \propto \phat^{-q}$ the steady state solution 
of eq. \ref{dce3.eq} is of the form $f(\phat) \propto \phat^{-(q+1)}$, $(q+1)=5$,
which implies a radio spectral index $\alpha = -1$.

This value of $\alpha$ is steeper than found in the
observational results,
although marginally consistent within  a 2 $\sigma$ margin. 
Table~\ref{tab:radio}
shows that in most of the regions of the radio source a 
rather flat spectrum $\alpha \approx -0.5$ is favoured. 
This value is in accord with what expected for a fresh distribution of CR
electrons accelerated at a strong shock in the test particle limit 
\cite{Drury83}. 
However, the spatial region sampled by the telescope beam is
larger than distance the CR electrons would propagate before
being affected by IC and synchrotron losses. 
In fact, the cooling time against IC losses of a relativistic 
electron emitting synchrotron radiation at frequency 
$\nu_{GHz}$ in units GHz
is $\tau_{IC} = 10^8 (B_\mu/\nu_{GHz})^{1/2}$ yr, where
$B_\mu$ is the magnetic field in $\mu$G. The distance
to which such electrons can be carried by the background flow
away from the acceleration region is L$=v_{flow}\tau_{IC}=
100 \; (v_{flow}/10^3 \, {\rm km\, s}^{-1}) \; (B_\mu/ \; 
\nu_{GHz})^{1/2}$ kpc. This distance is shorter than the 
telescope resolution length which at $z\simeq0.3$ and for the assumed 
cosmological model, corresponds to 400 kpc (for 1 arcmin).
Therefore, in accord with the above picture, 
we would expect the spectral index to be steeper than -0.5.

We can also compare the observed spectral index with the
``radio relic''  sources found in the peripheral regions of  
clusters which usually
have spectral indices $\alpha \lesssim -1$ and some times as
steep as $\alpha \lesssim -3$ \cite{1999NewA....4..141G,SRM01}.
In case of relics a very steep radio spectrum is explained in
a scenario in which  injection of fresh relativistic electrons in these
sources is believed to have ceased for a significant fraction
of their lifetime. This picture is different from the conditions expected in 
the radio filament
where ongoing {\it in situ} acceleration is likely to be taking place. Therefore
one would expect a spectral index flatter than the relic sources. 
However, it seems possible that the extended very steep spectrum source, detected only 
at 320 MHz
to the N-E of main filament, is in fact a region of 
relic emission, i.e., a remnant of past activity. 
In any case, a resolution of the spectral index discrepancy between the observed and
simulated values requires additional 
radio observations possibly over a much wider frequency range than 
explored here, which would allow the detailed spectral shape to be 
observed.     

The above assumptions on the acceleration efficiency
are consistent with a scenario in which
about 20-30\% of the shock ram pressure is converted into ionic CR
pressure and an  electrons to ions ratio at relativistic
energies, $R_{e/p}\sim 3-5 \times 10^{-2}$, in the range
allowed for the Galactic CRs \cite{muta87,mulletal95}.
With the total number of electrons fixed by
this choice of parameters for the acceleration efficiency,
the measured flux would imply a magnetic field strength
of order of $1\mu$G. This is consistent with values
inferred from Faraday rotation measures
for clusters of galaxies \cite{2001ApJ...547L.111C}. 
However, the
magnetic fields at accretion shocks have not been probed yet
and they could be weaker than assumed above.
In this respect, a detection of or an upper limit on
the inverse Compton HXR flux would provide an
important constraint on the magnetic field strength.
According to our estimate, even with an average magnetic field
as strong as 1-2$\mu$G the HXR flux should amount to about
$4\times 10^{-13}$ erg s$^{-1}$cm$^{-2}$.
Finally, the thermal bremsstrahlung emission produced by the
underlying collapsed structure generates a flux of
about 6 $\times 10^{-13}$ erg s$^{-1}$cm$^{-2}$,
corresponding to a $\sim 5$ keV cluster.
At the given redshift $z\sim 0.3$ this would imply a total luminosity
of $\sim  10^{44}$ erg s$^{-1}$. It is of interest to compare these values 
of thermal bremsstrahlung X-rays with those obtained from ROSAT data as
reported in \S~\ref{rosat} which are of the same order. 
We point out that these
values are associated with the whole emitting structure which, in the case
of Fig. \ref{fig13}, consists of a main collapsed object
and two smaller structures. These objects are lined up along a 
filamentary structure and are likely to merge later on
as their evolution progresses.

\section{Discussion, Conclusions and Outlook for Future} \label{conclude}

We have presented here the first  strong observational evidence for the
existence of large-scale shocks originating in the accretion flows of
intergalactic gas, as expected in theory of structure formation. The
presence of such shocks is inferred from the diffuse radio emission
extending over a physical scale of several Mpc, which intriguingly finds
a structural counterpart in  a filament of optical galaxies and in a
distribution of hot, clustered, thermal X-ray emitting gas.  The radio
emission appears to be the synchrotron radiation  from shock accelerated
cosmic-ray electrons and is  most likely associated with the
intergalactic gas as  a whole rather than with  individual sources.
This scenario is supported by  our analytical and numerical calculations
which indicate that  the above model can easily account for the detected
emission with conservative assumptions on the shock acceleration 
efficiency, and with a magnetic field strength of order  of a $\mu$G.

The implications of these observations are of great cosmological
interest, because they probe two important components of cosmic
environment: magnetic fields and cosmic-rays. They  indicate that
magnetic fields of significant strength are present not only in the
ICM but also in the diffuse inter-galactic medium, i.e., in the gas
that will be shocked as it accretes onto collapsing structures - the precursors
of virialized galaxy clusters.
In fact, these magnetic fields are responsible not only for providing the
scattering centers for the diffusive shock acceleration mechanism to
take place, but also for the synchrotron emission that we observed. No
significant magnetic field would be present in the post-shock region
of the flow, if it were not already present upstream of the shock.
Thus we have shown the first observational evidence for the existence
of magnetic fields in a super-cluster scale structure.
Since it is all but obvious how magnetic fields are amplified up to
such large values in cosmic filaments, these findings pose further
challenges to theoretical models.  At the same time, although 
particle acceleration at cosmic shocks is expected on theoretical
grounds \cite{Bell78,Blandford78}, theoreticians have been
``anxiously'' waiting for direct or indirect evidence that it really
occurs and at what level it could be linked to the origin of radio
relics (\eg \citeasnoun{mjkr01}, \citeasnoun{Ensslin98}). On 
the other hand, if diffusive
shock acceleration takes place with some efficiency during the
non-linear stage of large-scale structure formation, cosmic-ray ions
accumulating in the forming structure could become dynamically
important with interesting cosmological consequences \cite{mrkj01}.

The forming structure reported in this paper
could also be a source of high energy $\gamma$-ray photons
produced through CMBR inverse-Compton  
by relativistic electrons accelerated 
at large-scale structure formation shock-waves  up to  0.5~TeV 
\cite{Loeb2000}. Loeb \& Waxman claim that 
these electrons are responsible for the whole 
$\gamma$-ray background flux, although according to 
\citeasnoun{min02},
only a small fraction ($\sim$ 10-20\%) of such flux 
could be accounted for by inverse-Compton emission.
In any case, strong $\gamma$-emission is expected from accretion
flows around clusters [\citeasnoun{min02}, \citeasnoun{keshet02}, 
but see also \citeasnoun{totani2000}]. 

For this reason, the proto-cluster filament \astrobj{ZwCl 2341.1+0000} is
an ideal source  to check for $\gamma$-ray emission. 
Its diffuse radio luminosity of $\sim 40$ mJy can be
directly translated into an expected $\gamma$-ray flux above energy
$E_\gamma$ of $F_\gamma(>E_\gamma) = 10^{-10}\,{\rm photons\,\,
cm^{-2}\,s^{-1}}\, (B/\mu{\rm G})^{-2}\,(E_\gamma/{100 \,\rm
MeV})^{-1}$. For comparison - the sensitivities of the EGRET and
upcoming GLAST surveys are a few $10^{-8}$ and $10^{-9}\,{\rm
photons\,\, cm^{-2}\,s^{-1}}$ respectively, demonstrating that future
$\gamma$-ray facilities might become sensitive enough to address
these issues. 

Thus, the system \astrobj{ZwCl 2341.1+0000}  provides an interesting cosmic laboratory
where issues related to several high energy astrophysical phenomena can be 
observationally explored.
Further spectroscopic redshift mesurements and deep imaging of a wider region,  in
association of sensitive X-ray and $\gamma$-ray
observations of this structure will allow to study the energetic
processes in the  inter galactic medium in much greater detail.
We hope that they will be able to reveal further the physical mechanisms
that produce the relativistic electron population seen here in the
radio-synchrotron emission and will probe the  distribution of 
dark matter relative to the luminous matter and the
cosmic-ray particles. The detections of similar objects and signatures
of shock waves would help to test our understanding of the intergalactic
weather \cite{1998ApJ...502..518Q,Miniati00,Burns98,Ensslinetal00}, 
which is driven by
the power of cosmological  structure formation flows.

\begin{ack}
We thank Hans B{\"o}hringer for kindly providing 
us with the ROSAT X-ray images and for
his many helpful comments. This research has made use of the NASA/IPAC Extragalactic 
Database (NED)
which is operated by the Jet Propulsion Laboratory, Caltech, under contract
with the National Aeronautics and Space Administration. We acknowledge the use
of National Radio Astronomy Observatory (NRAO) VLA telescope.
The National Radio Astronomy Observatory is a facility of the National Science
Foundation operated under cooperative agreement by Associated Universities, Inc.

\end{ack}

\newpage

\begin{figure}
\leftmargin=2pc
\begin{center}
\end{center}
\caption{The co-added $R$-band CCD image of \astrobj{ZwCl 2341.1+0000}.
The image shown is an about 12 \arcmin $\times$ 12 \arcmin \, 
($4 \times 4 \, h^{-1}_{50} \rm Mpc $) size
field with north on top and east to
the left. The four  galaxies for which the spectroscopic redshifts are 
available in SDSS  and positions are listed in Table~\ref{tab:sdss}  
are shown by arrows}
\label{fig1}
\end{figure}

\newpage

\begin{figure}
\leftmargin=2pc
\begin{center}
\end{center}
\caption{The $(V-I)$ vs. $I$ colour-magnitude diagram of all detected 
galaxies in the
field of \astrobj{ZwCl 2341.1+0000}. The small dots are the data points and
encircled dots represent the specific colour selected 
E/S0 galaxies (see \S \ref{photomet}).
Their well defined linear colour-magnitude sequence from the entire
field strongly supports
the existence of a large-scale structure.}
\label{fig2}
\end{figure}

\newpage

\begin{figure}
\leftmargin=2pc
\begin{center}
\end{center}
\caption{The $(R-I)$ vs. $I$ colour-magnitude diagram of all detected
galaxies in the
field of \astrobj{ZwCl 2341.1+0000}. The small dots are  the data points and
encircled dots represent the specific colour selected E/S0 
galaxies (see \S \ref{photomet}).
Their well defined linear colour-magnitude sequence from the 
entire field strongly supports
the existence of a large-scale structure.}
\label{fig3}
\end{figure}

\newpage

\begin{figure}
\leftmargin=2pc
\begin{center}
\end{center}
\caption{The $(V-I)$ vs. $(V-R)$ colour-colour diagram of all
detected galaxies. The dotted line shows the colour evolutionary track
for a model E/S0 galaxy, starting at z=0 and going upto z=1. The
redshift values are indicated along the track with open circles.
The other symbols represent galaxy colours at z=0.3 of various
Hubble types: E/S0 (cross), Sa (star), Sb (downwards triangle),
Sc (upwards triangle), Sd (square), Irr (diamond) .}
\label{fig4}
\end{figure}

\newpage

\begin{figure}
\leftmargin=2pc
\begin{center}
\end{center}
\caption{The $(R-I)$ vs. $(V-I)$ colour-colour diagram of all
detected galaxies. The dotted line shows the colour evolutionary track
for a model E/S0 galaxy, starting at z=0 and going upto z=1. The
redshift values are indicated along the track with open circles.
The other symbols represent galaxy colours at z=0.3 of various
Hubble types: E/S0 (cross), Sa (star), Sb (downwards triangle),
Sc (upwards triangle), Sd (square), Irr (diamond) .}
\label{fig5}
\end{figure}

\newpage

\begin{figure}
\leftmargin=2pc
\begin{center}
\end{center}
\caption{The VLA 1.4 GHz radio contour map, shown
superposed on the $R$-band CCD image of \astrobj{ZwCl 2341.1+0000}. The
contour levels are at 0.55 mJy/beam $\times$ [-6.4,-3.2,-1.6,1.6,3.2,6.4,12.8,25.6],
i.e. multiples of r.m.s. noise level of 0.55 mJy/beam and spaced in steps of 2.  
The beam size (1 arcmin FWHM) is shown at the
top right corner.}
\label{fig6}
\end{figure}

\newpage
\begin{figure}
\leftmargin=2pc
\begin{center}
\end{center}
\caption{The high resolution VLA `FIRST' survey radio contour map
at 1.4 GHz. Shown in the
background is the Palomar Digitized Sky Survey
(DSS-2) E-plate (red) image. The contour levels are
at 1 mJy/beam $\times$ [-0.4, 0.4, 0.8, 1.60, 2.50] and 
the rms of noise is 0.15 mJy/beam.
The beam size  ($\rm {6.4 \ arcsec \times 5.4 \ arcsec}$ FWHM) is shown at the
top right corner.}
\label{fig7}
\end{figure}

\newpage

\begin{figure}
\leftmargin=2pc
\begin{center}
\end{center}
\caption{VLA 1.4 GHz `FIRST' radio map of the diffuse radio
emission associated with the radio source 
designated NS in the northern half of the
radio filament (\S~\ref{radobs}). }
\label{fig8}
\end{figure}

\newpage

\begin{figure}
\leftmargin=2pc
\begin{center}
\end{center}
\caption{The VLA 320 MHz radio contour map, shown
superposed on the $R$-band CCD image of \astrobj{ZwCl 2341.1+0000}. The
contour levels are at 2.5 mJy/beam $\times$ [-6.4,-3.2,-1.6,1.6,3.2,6.4,12.8,25.6],
i.e. multiples of r.m.s. noise level of 2.5 mJy/beam and spaced in steps of 2.
The beam size (1.8 arcmin FWHM) is shown at the
top right corner.}

\label{fig9}
\end{figure}

\newpage


\begin{figure}
\leftmargin=2pc
\begin{center}
\end{center}
\caption{Detailed radio contour map of the extended region of
diffuse emission mapped by VLA at 320 MHz. In the
background is the Palomar Digitized Sky Survey
(DSS-2) E-plate (red) image. The contour levels are
at 2.5 mJy/beam $\times$ [-4.5,-3.2,-2.2,-1.6,1.6,2.2,3.2,4.5,6.4,9,12.8],
i.e. multiples of r.m.s. noise level of 2.5 mJy/beam and spaced in steps
of $\sqrt 2$.
The beam size (1.8 arcmin FWHM) is shown at the
top right corner.}
\label{fig10}
\end{figure}

\newpage


\begin{figure}
\leftmargin=2pc
\begin{center}
\caption{The ROSAT X-ray detection in the region of \astrobj{ZwCl 2341.1+0000},
from the ROSAT (PSPC) All-sky survey archival data.
We have used data in the 0.5-2.0 keV energy range and have smoothed the
original data with a Gaussian of $\sigma$ of about 1 arcminute. The X-ray
contours at the levels (4,5,6,7,8,9) $\times \ 5\times 10^{-3}$ counts/s, i.e.
in multiples of the r.m.s. noise level of $5\times 10^{-3}$ counts/s,
are shown superposed on the Palomar Digitized Sky Survey
(DSS-2) E-plate (red) image. } 
\label{fig11}
\end{center}
\end{figure}

\newpage
\begin{figure}
\leftmargin=2pc
\begin{center}
\end{center}
\caption{A superposition of ROSAT X-ray data (in contours) over
the VLA 320 MHz radio map shown as a B/W photograph. The 
 X-ray contour values are the same as defined in Fig. 11.}
\label{fig12}
\end{figure}
\newpage

\begin{figure}
\leftmargin=2pc
\begin{center}
\end{center}
\caption{Images from hydro + N-body structure formation
simulation. Each panel is a 2-D projection $ \approx $ 
10 Mpc $ \times $ 10 Mpc 
in size  showing the appearence of a forming structure at the
redshift z=0.3. To better compare with the observational radio images, the
value of radiation emissions integrated along  line of sight in synthetic 
images have been averaged over $3 \times 3$ pixels
(i.e. $ \approx  1^{\prime} \times 1^{\prime}$ ). 
Top left: Radio emission map at 1.4 GHz. The log of the flux density is shown
in units of `$R_{e/p}^{-1}$ Jy pxl$^{-1}$'. 
Bottom left: The
iso-contours
of the radio emission at 1.4 GHz, ranging from $10^{-5.8}$ to $10^{-4.6}$
and separated by constant factor of $10^{0.15}$.
Top right:
X-rays from thermal bremsstrahlung in the energy range 0.5-2.0 keV. The 
log of the flux is shown in units of
`erg s$^{-1}$ cm$^{-2}$ pxl$^{-1}$'.
Bottom right: Inverse Compton hard X-ray emission at 50 keV. The  
log of the monochromatic flux is shown in units 
`$R_{e/p}^{-1} 10^{-23}$ erg s$^{-1}$ cm$^{-2}$ Hz$^{-1}$
pxl$^{-1}$'. In these radio or X-ray units, 1 pxl is 
about $\rm 0.34 \times 0.34 \,\, arcmin^{2}$. 
The factor $R_{e/p}$ is the ratio of number of 
cosmic-ray electrons to protons at relativistic energies, on which the
non-thermal radiation flux is dependent (see \S~\ref{numerical}).} 

\label{fig13}
\end{figure}

\end{document}